\documentclass[%
 aip,apl,
 amsmath,amssymb,
 reprint,natbib,%
 nofootinbib
]{revtex4-1}

\usepackage{graphicx}
\usepackage{dcolumn}
\usepackage{bm}
\usepackage[utf8]{inputenc}
\usepackage[T1]{fontenc}
\usepackage{mathptmx}
\usepackage{siunitx}
\usepackage{xr}

\makeatletter
\newcommand*{\addFileDependency}[1]{
  \typeout{(#1)}
  \@addtofilelist{#1}
  \IfFileExists{#1}{}{\typeout{No file #1.}}
}
\makeatother

\newcommand*{\myexternaldocument}[1]{%
    \externaldocument{#1}%
    \addFileDependency{#1.tex}%
    \addFileDependency{#1.aux}%
}

\myexternaldocument{Supplements/SupplementalMaterial}

\begin{document}

\preprint{AIP/123-QED}

\title[Sample title]{Hybrid magnetization dynamics in $\text{Cu}_\text{2}\text{OSeO}_\text{3}$/NiFe heterostructures}

\author{Carolina Lüthi$^*$}
\affiliation{Walther-Meißner-Institut, Bayerische Akademie der Wissenschaften, Garching, Germany}
\affiliation{Physics Department, Technical University of Munich, Garching, Germany}

\author{Luis Flacke$^*$}
\affiliation{Walther-Meißner-Institut, Bayerische Akademie der Wissenschaften, Garching, Germany}
\affiliation{Physics Department, Technical University of Munich, Garching, Germany}

\author{Aisha Aqeel}
\affiliation{Physics Department, Technical University of Munich, Garching, Germany}
\affiliation{Munich Center for Quantum Science and Technology (MCQST), Munich, Germany}

\author{Akashdeep Kamra}
\affiliation{Condensed Matter Physics Center (IFIMAC) and Departamento de F\'{i}sica Te\'{o}rica de la Materia Condensada, Universidad Aut\'{o}noma de Madrid, Madrid, Spain}

\author{Rudolf Gross}
\affiliation{Walther-Meißner-Institut, Bayerische Akademie der Wissenschaften, Garching, Germany}
\affiliation{Physics Department, Technical University of Munich, Garching, Germany}
\affiliation{Munich Center for Quantum Science and Technology (MCQST), Munich, Germany}

\author{Christian Back}
\affiliation{Physics Department, Technical University of Munich, Garching, Germany}
\affiliation{Munich Center for Quantum Science and Technology (MCQST), Munich, Germany}

\author{Mathias Weiler}
\email{weiler@physik.uni-kl.de}
\affiliation{Walther-Meißner-Institut, Bayerische Akademie der Wissenschaften, Garching, Germany}
\affiliation{Physics Department, Technical University of Munich, Garching, Germany}
\affiliation{Fachbereich Physik and Landesforschungszentrum OPTIMAS, Technical University of Kaiserslautern, Kaiserslautern, Germany}

\date{\today}

\def\thefootnote{*}\footnotetext{These authors contributed equally to this work}

\begin{abstract}
We investigate the coupled magnetization dynamics in heterostructures of a single crystal of the chiral magnet $\mathrm{Cu_2OSeO_3}$ (CSO) and a polycrystalline ferromagnet $\mathrm{NiFe}$ (Py) thin film using broadband ferromagnetic resonance (FMR) at cryogenic temperatures. We observe the excitation of a hybrid mode (HM) below the helimagnetic transition temperature of CSO. This HM is attributed to the spin dynamics at the CSO/Py interface. 
We study the HM by measuring its resonance frequencies for in plane rotations of the external magnetic field.
We find that the HM exhibits dominantly four-fold anisotropy, in contrast to the FMR of CSO and Py.
\end{abstract}

\maketitle

Chiral magnets exhibit non-collinear spin structures such as spin helices and magnetic skyrmions below their critical temperature $T_c$ and critical field $H_{c2}$ \cite{Stasinopoulos.2017}. Skyrmions are topologically protected non-coplanar magnetization configurations that can behave as particle-like objects. Furthermore, they are small yet stable making them suitable to become the carriers of information in future devices \cite{Back.2020, Okuyama.2019,Legrand.2017,Woo.2016, Jiang.2015, Fert.2013, Sampaio.2013, Jonietz.2010,Yu.2012,Yu.2010,Muhlbauer.2009}.
The non-collinear spin structure of chiral magnets gives rise to intriguing magnetization dynamics, in particular in their skyrmion lattice phase \cite{Schwarze.2015}. The recently discovered low-temperature skyrmion phase \cite{Chacon.2018} leads to additional striking spin dynamical signatures \cite{Aqeel.2021} in the low-damping \cite{Stasinopoulos.2017.two} chiral magnet CSO. The periodicity of the magnetic lattice leads to naturally formed magnonic crystals \cite{Weiler.2017} and the skyrmion eigenmodes can be coupled to photonic resonators with high cooperativity \cite{Liensberger.2021}. The chiral properties of skyrmions give rise to non-reciprocal spin-wave dynamics \cite{Seki.2020}. Thus, the emerging field of spin dynamics of chiral magnets has already revealed important fundamental insights with perspectives for practical applications. 

In topologically trivial magnets, the now well studied coupling between multiple magnetic layers \cite{Bruno.1995, Crew.2003, GonzalezChavez.2013, Heinrich.2003, McMichael.1998, Schafer.2012, Woltersdorf.2007} resulted in the discovery of some of the most technologically relevant effects such as tunneling magnetoresistance \cite{Julliere.1975} or giant magnetoresistance \cite{Baibich.1988, Binasch.1989}. Heterostructures of topologically trivial magnets can exhibit coupled spin dynamics that can lead to, e.g., excitation of nanoscale spin waves \cite{Klingler.2018}. Much less is known about spin dynamics in heterostructures of collinear and chiral magnets. The coupling between distinct order parameters across interfaces has explained important phenomena such as proximity effects, exchange bias or exchange spring-induced hard magnets \cite{Hellman.2017}. However, the studies of excitations in chiral magnets are so far limited to a single magnetically ordered constituent \cite{Takeuchi.2019}. Even though the formation of novel topological order at the chiral magnet/ferromagnet interface was predicted by theory it has not yet been observed in experiment \cite{Kawaguchi.2016}.

In this work, we investigate the hybrid magnetization dynamics of heterosturctures of thin film metallic ferromagnets and bulk chiral magnets, in this case the ferrimagnetic insulator $\mathrm{Cu_2OSeO_3}$ (CSO).
To study the magnetization dynamics of the chiral magnet/ferromagnet heterostructures in the GHz frequency regime we use broadband ferromagnetic resonance spectroscopy. We experimentally determine and phenomenologically model the resonance frequencies in such heterostructures. Thereby, we find that a hybrid mode of the chiral magnet/ferromagnet heterostructure is excited, which we attribute to the spin dynamics at the interface of the two magnetic layers.
\begin{figure}[htb!]
	\includegraphics[width=80mm]{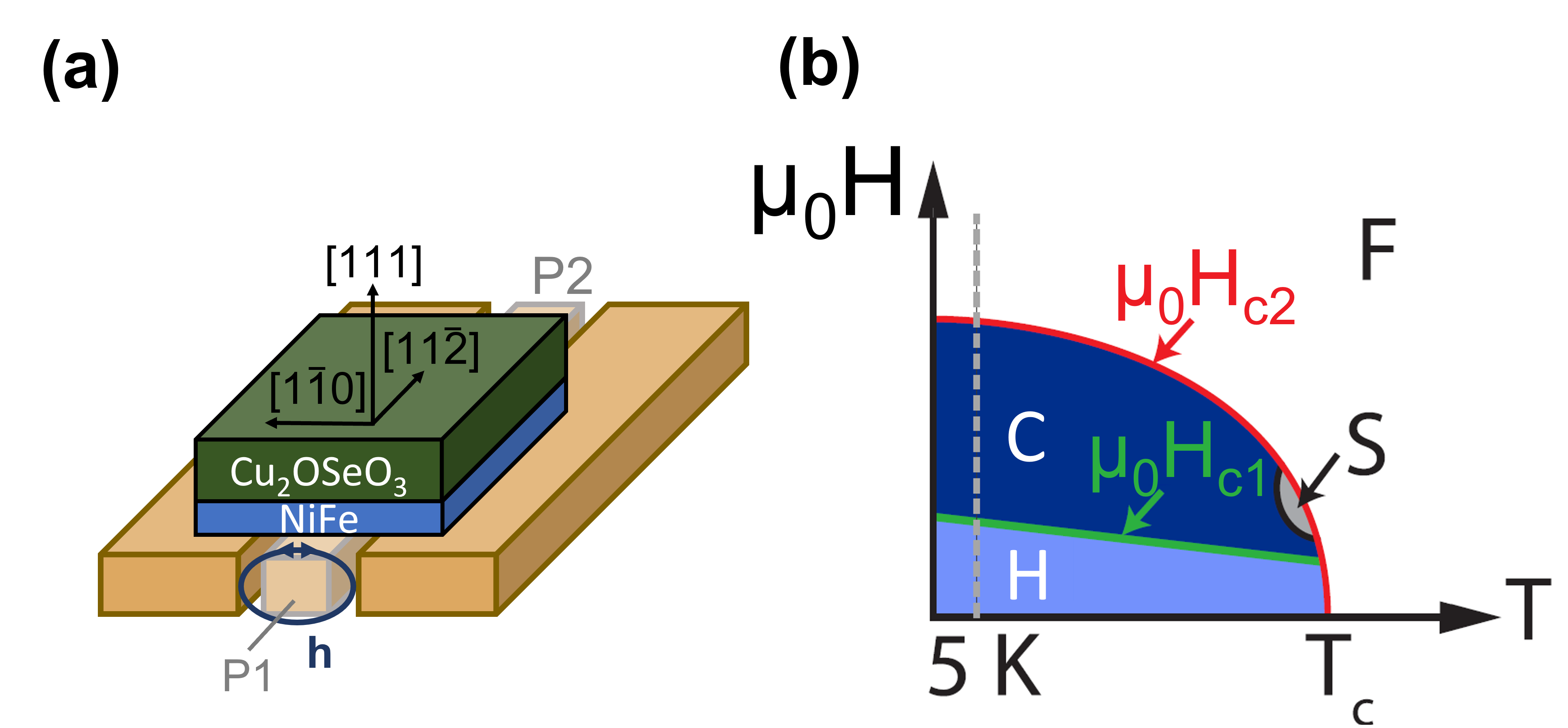} 
	\caption{\label{fig_1} (a) Experimental setup: The (111)-oriented $\mathrm{Cu_2OSeO_3}/\mathrm{NiFe}$ heterostructure is placed on the CPW, which generates a magnetic field ($\mathbf{h}$) within the sample due to the application of an ac current flowing from port 1 (P1) to port 2 (P2) of the center conductor. (b) Schematic phase diagram of the $\mathrm{Cu_2OSeO_3}$ crystal defining the helimagnetic transition temperature $T_c = 58.2$~K \cite{Qian.2016} as well as the critical fields $\mu_0H_{c1}$ and $\mu_0H_{c2}$ (H: helical state, C: conical state, S: skyrmionic state, F: ferrimagnetic state). The vertical dashed line indicates the phases of the CSO at 5~K in dependence of the external field.}
\end{figure}
We investigate a $\mathrm{CSO/Ni_{80}Fe_{20}}$ (CSO/Py) sample, where the Py thin film has a thickness of 40 nm. The CSO crystal is (111)-oriented and cut to a cuboid shape with dimensions $L_x = 2.5$ mm, $L_y = 1.5$ mm, and $L_z = 0.8$ mm. It was grown by a chemical vapor transport method \cite{Aqeel.2021}(for details on the crystal orientation of the CSO and the sample preparation see Supplemental Material \ref{sec:crystal} and \ref{sec:sample} \cite{.}). We place the CSO/Py hybrid on top of a coplanar waveguide (CPW) with a center conductor width of $w = 127 \mu$m as shown in Fig.~\ref{fig_1}~(a). The CPW is connected to two ports P1 and P2 of a vector network analyzer (VNA), which measures the change of transmission from P1 to P2 defined as the complex transmission parameter $S_{21}$ as a function of frequency $f$ and external magnetic field $\mu_0H$ at a fixed microwave power of 1 mW (0 dBm). We then place the CPW/CSO/Py assembly into the variable temperature insert of a superconducting 3D-vector magnet. By applying a static external magnetic field $\mu_0 \mathbf{H}$ in the plane of the Py thin film and setting the temperature to 5~K we can access the helical $(H)$, conical $(C)$, and ferrimagnetic $(F)$ phases of the CSO as schematically depicted in Fig.~\ref{fig_1}~(b).
\begin{figure}[htb!]
	\includegraphics[width=85mm]{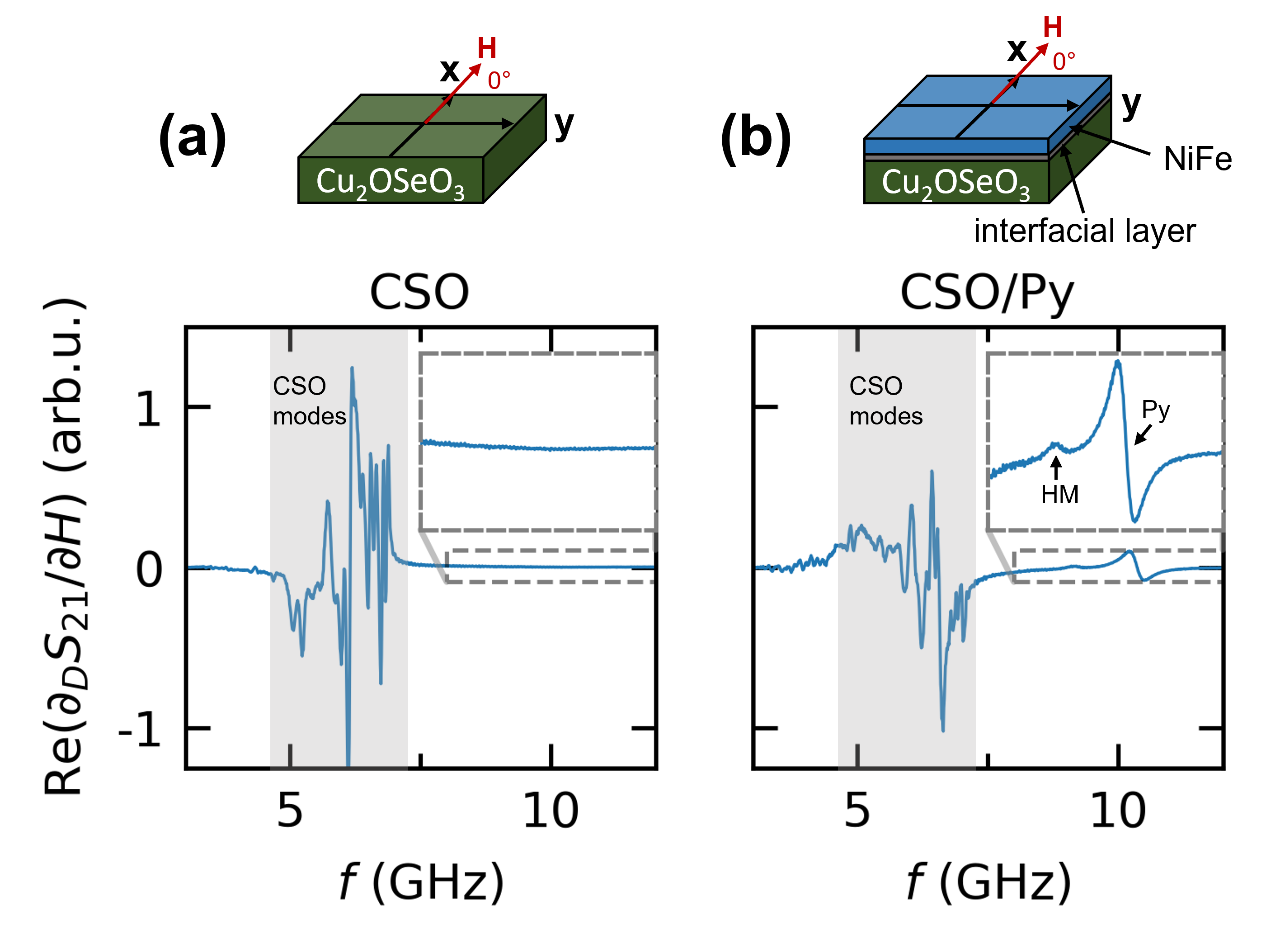}
	\caption{\label{fig_2} Measured broadband ferromagnetic resonance spectrum of the CSO/Py sample at 5~K as a function of the frequency $f$. The fixed external magnetic field $\mu_0\mathbf{H} = 120$~mT is applied along the $\mathbf{x}$-axis ($\phi_H = 0^{\circ}$) as indicated in the sample sketches. (a) The CSO crystal faces the CPW. Thus, only ferrimagnetic CSO modes apppear (grey marked frequency range). (b) The Py thin film faces the CPW. In addition to the CSO modes the Py FMR mode appears at high frequencies as well as a hybrid mode (HM) at medium frequencies (inset). This HM is attributed to the spin dynamics at the interface of the CSO/Py sample, indicated as an interfacial layer in the sample sketch.}
\end{figure}

As a reference measurement, we first place the CSO/Py hybrid sample on the CPW with the Py facing away from the CPW. Due to the large thickness of the CSO layer, the oscillating magnetic field generated by the CPW does not reach the Py layer. In this way, we only excite the magnetization in CSO itself with no influence of the Py layer. We apply a fixed external magnetic field $\mu_0 H~=~120$~mT along the $\mathbf{x}$-axis ($\phi_H~=~0^{\circ}$) at 5~K, as illustrated schematically in the top panel of Fig.~\ref{fig_2}~(a). For this temperature and external magnetic field strength the CSO magnetization is in the field polarized phase. To correct for the microwave background of the complex transmission parameter $S_{21}$ we use the derivative divide method \cite{MaierFlaig.2018}, to obtain the field derivative of the complex transmission parameter $\frac{\mathrm{\partial} S_{\mathrm{21}}}{\mathrm{\partial}H}$ devided by $S_{21}$.
On the bottom panel of Fig.~\ref{fig_2}~(a) $\mathrm{Re}(\mathrm{\partial}_{\mathrm{D}} S_{\mathrm{21}}/\mathrm{\partial}\mathrm{H})$ for the CPW/CSO/Py assembly at 5~K is shown as a function of the frequency $f$. In the grey marked frequency range several resonances appear. These are attributed to the excitation of magnetostatic modes of the cuboid-shaped CSO crystal. In the frequency range 7~GHz~<~$f$~<~12~GHz no additional modes are observed (inset).
After determining the response of the isolated CSO magnetization dynamics, we place the CSO/Py hybrid on the CPW with the Py facing the CPW and again apply a fixed magnetic field $\mu_0 H = 120$~mT along the $\mathbf{x}$-axis as schematically shown in the top panel of Fig.~\ref{fig_2}~(b). Now, the field generated by the CPW interacts with the Py layer as well as the CSO as the Py layer is a thin film.
In the bottom panel of Fig.~\ref{fig_2}~(b), $\mathrm{Re}(\mathrm{\partial}_{\mathrm{D}} S_{\mathrm{21}}/\mathrm{\partial}\mathrm{H})$ at 5~K is shown for the CPW/Py/CSO assembly as a function of the frequency $f$.  In addition to the resonance lines of CSO (grey marked frequency range) also the Py FMR line appears close to 10~GHz as expected. The change in the CSO mode spectrum is attributed to the presence of the metallic film and concomitant shielding of the microwave field in the bulk of CSO. Furthermore, we observe an additional medium frequency mode, which is shifted by about 1~GHz to lower frequencies than the Py FMR mode. This hybrid mode (HM) is also observed in the CSO/Py hybrid in the conical phase of CSO (see Fig.~\ref{fig:conical} in Supplemental Material \cite{.}). For the dependence of the HM on the magnitude of the external magnetic field see Supplemental Material \ref{sec:colorplot} \cite{.}. We attribute the appearance of this additional mode to the spin dynamics at the CSO/Py interface. Thus, in a simplified macrospin picture, we may treat our bilayer as a trilayer with a new interfacial layer inheriting properties from both sides. The HM is then modelled as a result of macrospin dynamics of the interlayer as discussed in the following.
\begin{figure}
	\includegraphics[width=85mm]{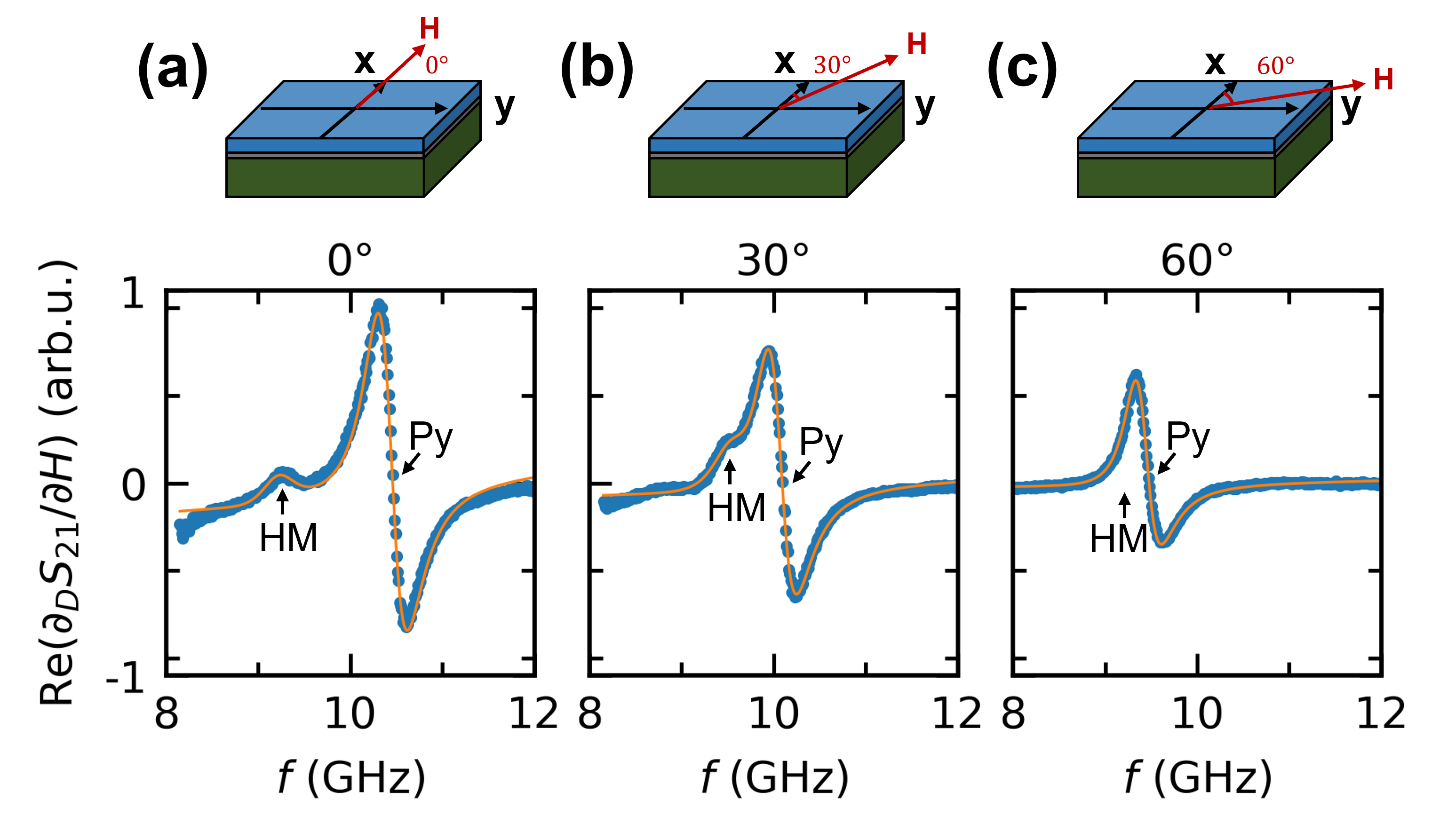}
	\caption{Fits of the frequency spectrum for a fixed magnitude of the external field $\mu_0\mathrm{H}~=~120$~mT at 5~K. The magnetic field direction is rotated in the Py thin film plane and defined by the angle $\phi_H$ with respect to the $\mathbf{x}$-axis as indicated in the sketches of the sample. The blue dots correspond to the measurement data and the orange line to the fit. (a) $\phi_H~=~0^{\circ}$. The peak of the HM at 9.2~GHz and the Py FMR peak-dip at 10.4~GHz are clearly separable. (b) $\phi_H~=~30^{\circ}$. The two resonance frequencies of the Py mode and the HM are less separated. (c) $\phi_H = 60^{\circ}$. The two resonance frequencies of the Py mode and the HM are not separable any more.}
	\label{fig_3}
\end{figure}
 \begin{figure*}
	\includegraphics[width=175mm]{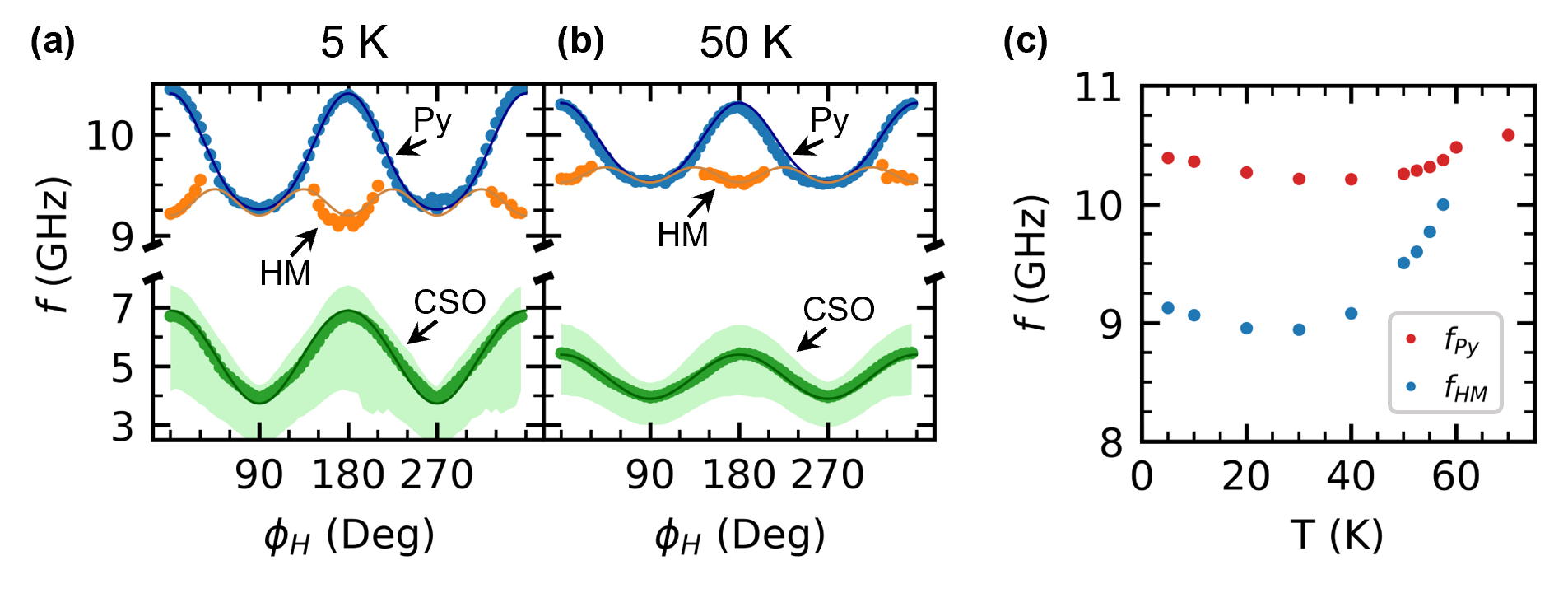}
	\caption{\label{fig_4}(a) and (b) Dependence of the Py (blue dots), CSO (green dots), and HM (orange dots) resonance frequencies on the external magnetic field direction for a 360° rotation in the Py thin film plane and a fixed magnitude $\mathrm{\mu_0 H}~=~120$~mT at 5~K (a) and 50~K (b). The green shaded range around the CSO resonance frequencies indicates the frequency distribution of the magnetostatic CSO modes. The fit errors are smaller than the symbol size. The solid lines show the simulation result according to Eq. (\ref{LLGi}) of the Py (blue line), CSO (green line), and HM (orange line) resonance frequencies which are in good agreement with the measurement data. (c) Fit result of the Py (red dots) and HM (blue dots) resonance frequencies in dependence of the temperature with fixed external field  $\mathrm{\mu_0 H}~=~120$~mT applied along the $\mathbf{x}$-axis ($\phi_H~=~0$). Above the critical temperature $T_c = 58.2$~K of the CSO the HM mode vanishes.}
\end{figure*}

To investigate the dependence of the HM on the direction of the external magnetic field we apply a field with a fixed magnitude of $\mu_0H~=~120$~mT and rotate the field direction by $360^{\circ}$ in the Py film plane. For a quantitative analysis of the HM we simultaneously fit the Py FMR peak and the HM peak in the frequency domain of the transmission parameter $S_{21}$ for each fixed external field direction (for detailed informations on the fit model see Supplemental Material \ref{sec:fitting} \cite{.}).
In Fig.~\ref{fig_3} three exemplary fits of the frequency spectrum are shown for a fixed magnitude of the field $\mu_0 H = 120$~mT and for different directions under which it was applied.
In Fig.~\ref{fig_3}~(a) the external field is applied in the Py thin film plane along the $\mathbf{x}$-axis as schematically depicted at the top of Fig.~\ref{fig_3}~(a). We find the peak of the hybrid mode at 9.2~GHz, which has a small amplitude compared to the Py FMR peak-dip at 10.4~GHz.
In Fig.~\ref{fig_3}~(b) we show the fit result for the field applied under an angle $\phi_H~=~30^{\circ}$ with respect to the $\mathbf{x}$-axis as depicted at the top of Fig.~\ref{fig_3}~(b). Now, the two resonance frequencies of the Py mode and the HM are less separated than in Fig.~\ref{fig_3}~(a), as the hybrid mode moved to higher frequencies and the Py mode to lower frequencies. 
In Fig.~\ref{fig_3}~(c) the field is applied under an angle $\phi_H~=~60^{\circ}$. The two resonance frequencies of the Py mode and the HM are not separable anymore as the HM resonance frequency is presumably superimposed on the Py resonance frequency. Furthermore, the amplitude of the FMR signal decreases in Fig.~\ref{fig_3}~(b)~and~(c). As the applied field shifts away from the $\mathbf{x}$-axis, so does the equilibrium magnetization. As a result, the CPW oscillating field component transverse to the static magnetization also diminishes. This results in a reduction of the recorded FMR signal amplitude.

To better understand the behaviour of the spin dynamics in the sample in dependence of the external magnetic field and temperature we fit the resonance frequencies of each layer for a full rotation of the external field applied in the Py thin film plane and we evaluate the dependence of the HM resonance frequency on temperature by fitting its resonance frequency for a fixed field direction at different temperatures.
In Fig.~\ref{fig_4}~(a) the fitted resonance frequencies at 5~K of the Py FMR mode (blue dots), the CSO modes (green shading indicates the frequency range over which the CSO magnetostatic modes are observed) and the hybrid mode (orange dots) are plotted against the angle $\phi_H$. In the fit model, we only consider the CSO mode with the strongest amplitude (green dots). Furthermore, for $\phi_H \simeq$~90° we are not able to fit the HM, as it either vanishes or merges with the Py FMR mode.
Due to the demagnetization field in the CSO crystal the CSO resonance frequencies show a cosine like dependence on the angle under which the external field is applied. Furthermore, we observe an uniaxial anisotropy in the Py resonance frequencies. This uniaxial anisotropy is strongest for 5~K and weakens for increasing temperatures (See Fig. \ref{fig_4}~(b) and Supplemental Material \ref{sec:uniaxialani} \cite{.}). Additionally, a broadening of the Py resonance minima and a narrowing of the respective peaks is observed. The HM shows a striking feature with its angle dependence being inverted with respect to the CSO and Py modes. Its presence and frequency difference with respect to the Py mode at $\phi_H = 0$ vanishes when surpassing the CSO ordering temperature as shown in Fig. \ref{fig_4}~(c).

The angle and temperature dependence of the observed modes reveals a complex interaction within the heterostructure. The following explanation attempts expose the non-trivial nature of the system by leading to major discrepancies when assuming simple models with isotropic interaction parameters.

First, we start by noting the effect of CSO on the system. By increasing the temperature a reduced uniaxial anisotropy not only of the CSO but also of the Py and HM mode becomes evident (see Fig.~\ref{fig_4}~(a) and (b) and Supplemental Material \ref{sec:uniaxialani} \cite{.}). This reduction agrees with a decreased spin ordering in CSO at higher temperatures, such that the effect of magnetic stray field of the bulk CSO crystal on the Py layer can be assumed to be the dominant effect seen by the uniaxial anisotropy in the Py mode (see calculation of the anisotropy field according to Ref. \onlinecite{EngelHerbert.2005} in the Supplemental Material \ref{sec:uniaxialani}).

It also supports our inference that CSO plays the main role in determining the frequency of the HM. In the following, we argue that the HM is indeed the result of a coupling of the multilayer system. Figure~\ref{fig_4}~(c) visualizes the dependence of the Py and the HM resonance frequencies on temperature where the external field has a magnitude of 120~mT and is applied under an angle $\phi_H~=~0$°. With increasing temperature the HM mode approaches the Py mode until it vanishes above 58~K. Thus, above the critical temperature of CSO, the coupling of the CSO and Py dynamics vanishes and the HM cannot be excited any more. This demonstrates that the existence of the HM is a direct consequence of the magnetic ordering of CSO. We thus exclude non-uniformity of the Py layer as its origin. Its inverted angular dependence further substantiates this conclusion as otherwise equal symmetry with respect to the Py and CSO modes can be expected.

Assuming exchange interaction and spin torques at the interface similar as in Ref.~\onlinecite{Klingler.2018} and including such terms in the Landau-Lifshitz-Gilbert (LLG) equation would result in four solutions for resonance frequencies. Namely, the uncoupled CSO and Py layer, as well as two coupled modes. It might be possible that the second coupled mode is hidden in the series of magnetostatic modes of the CSO. But in this case, again, the solutions for the hybrid modes would only lead to a constant shift of the HM frequencies compared to the Py resonance frequencies without an angle dependent gap between the Py and the HM resonance frequencies. Nonetheless, if the exchange interaction was to be anisotropic, the additional degree of freedom could result in the observed symmetry of the HM.

CSO is known for its intrinsic symmetry breaking and the existence of Dzyaloshinskii-Moryia interaction (DMI). The Py thin film experiences a symmetry breaking at the interface with CSO and a residual effect of the DMI could influence the magnetization dynamics of Py at the interface resulting in a virtual additional layer. As the DMI favors a perpendicular alignment the excitation of magnetization is favoured and its interaction would reduce the resonance frequency due to reduced total energy~\cite{Garst.2017}. However, also in this case, an anisotropic interaction parameter has to be assumed to explain the angle dependence as otherwise a constant frequency shift is to be expected again~\cite{Garst.2017}. On the other hand, if we consider the HM to possess a nonzero inplane wavevector, it would not be influenced by the uniaxial shape anisotropy that appears to dominate the Py and CSO resonance frequencies. The angle-dependence of the HM is still not fully captured in that case, and an additional origin needs to be found for it.

Another observation which makes a clear interpretation even more difficult is the broadening/narrowing of the Py resonance minima/maxima. Such a behaviour is usually seen in material systems with a cubic, i.e., four-fold symmetry superimposed with the previously described uniaxial anisotropy. This contradicts the three-fold symmetry that is expected from the (111)-orientation of the CSO crystal. Naively, a cubic magnetocrystalline anisotropy in the Py layer could explain the observation~\cite{Liensberger.2019} . Yet it is not clear why such a crystalline structure should have developed when considering either the (111)-oriented substrate or the continuous rotation of the sample during the deposition process.

The chiral magnet/ferromagnet heterostructure proves to be a highly complex material system raising further questions about anisotropic interaction parameters and fundamental symmetry manifestations that go beyond the scope of this work.
To phenomenologically model the data in Fig.~\ref{fig_4}~(a)~and~(b) we use a Landau-Lifshitz-Gilbert (LLG) approach, where we treat the magnetization dynamics in the sample as a three-layer system consisting of the uncoupled Py and CSO layer as well as an interfacial layer as shown at the top of Fig.~\ref{fig_2}~(b).
Thereby, the magnetization dynamics of each of these three layers are treated as macrospins. 
The equation of motion for magnetization $\mathbf{M}_i$ then reads
\begin{eqnarray}
	\frac{\mathrm{d}\mathbf{M}_i}{\mathrm{d}t} =&& - \gamma_i \mathbf{M}_i \times \mu_0 \mathbf{H}_{\mathrm{eff}_i}(\phi_H) + \frac{\alpha_i}{M_{s_i}}\mathbf{M}_i \times \frac{\mathrm{d}\mathbf{M}_i}{\mathrm{d}t}.
	\label{LLGi}
\end{eqnarray}
Here, $\gamma_i$ is the gyromagnetic ratio and $\alpha_i$ the Gilbert damping parameter of layer $i$. For all three magnetic layers the effective field $\mathbf{H}_{\mathrm{eff}_i}$ accounts for the external field, the driving field, and the demagnetization field.
In case of the Py layer $\mathbf{H}_{\mathrm{eff}}$ additionally includes a phenomenological uniaxial magnetic anisotropy, which is expected due to the demagnetization fields of the CSO and can be regarded effectively as a shape anisotropy of the CSO. However, also a finite contribution due to the trapped flux in the superconducting coils is to be expected (See Supplemental Material \ref{sec:uniaxialani}) \cite{.}.
For both the Py layer and the interfacial layer we also include a phenomenological anisotropy with a four-fold symmetry in accordance with our experimental observation. We note, again, that the four-fold symmetry does not correspond to the  cubic anisotropy of CSO, because the projection of the cubic anisotropy to the (111) plane would result in a three-fold symmetry. The effective field of layer $i$ is given by the Eqs.~(\ref{eq:Heffpy}),~(\ref{eq:HeffHM}),~and~(\ref{eq:HeffCSO}) in Supplemental Material \ref{sec:simulation} \cite{.}.
In Fig.~\ref{fig_4}~(a)~and~(b) the Py (blue line),  HM (orange line), and CSO (green line) resonance frequencies simulated with Eq.~(\ref{LLGi}) are in good agreement with the measurement data (symbols). 

In conclusion, we investigated the coupled magnetization dynamics in a CSO/Py heterostructure by broadband ferromagnetic resonance experiments at cryogenic temperatures.
We found that a hybrid mode at the CSO/Py interface is excited. While a microscopic picture for the formation of the HM is so far missing, our experimental findings pave the way for future experiments on coupled spin dynamics in topologically non trivial magnetic bilayer systems, which have recently attracted great interest from the viewpoint of possible applications in high-performance memory devices \cite{Takeuchi.2019, Soumyanarayanan.2016,Fert.2013}.

\begin{acknowledgments}
	We gratefully acknowledge the financial support by the Deutsche Forschungsgemeinschaft (DFG, German Research Foundation) via WE 5386/5-1 and BA 2181/19-1 and financial support from the Spanish Ministry for Science and Innovation -- AEI Grant CEX2018-000805-M (through the ``Maria de Maeztu'' Programme for Units of Excellence in R\&D).
	We would also like to thank M. Müller for support in carrying out some of the measurements.	
\end{acknowledgments}

\section*{Data Availability Statement}
The data that support the findings of this study are available from the corresponding author upon reasonable request.

\bibliography{Quellen}

\newpage

\section*{Supplementary information for the article \\
	Hybrid magnetization dynamics in $\text{Cu}_\text{2}\text{OSeO}_\text{3}$/NiFe heterostructures}

\maketitle
\section{CSO crystal orientation}
\label{sec:crystal}

\begin{figure}[htb!]
	\includegraphics[width=40mm]{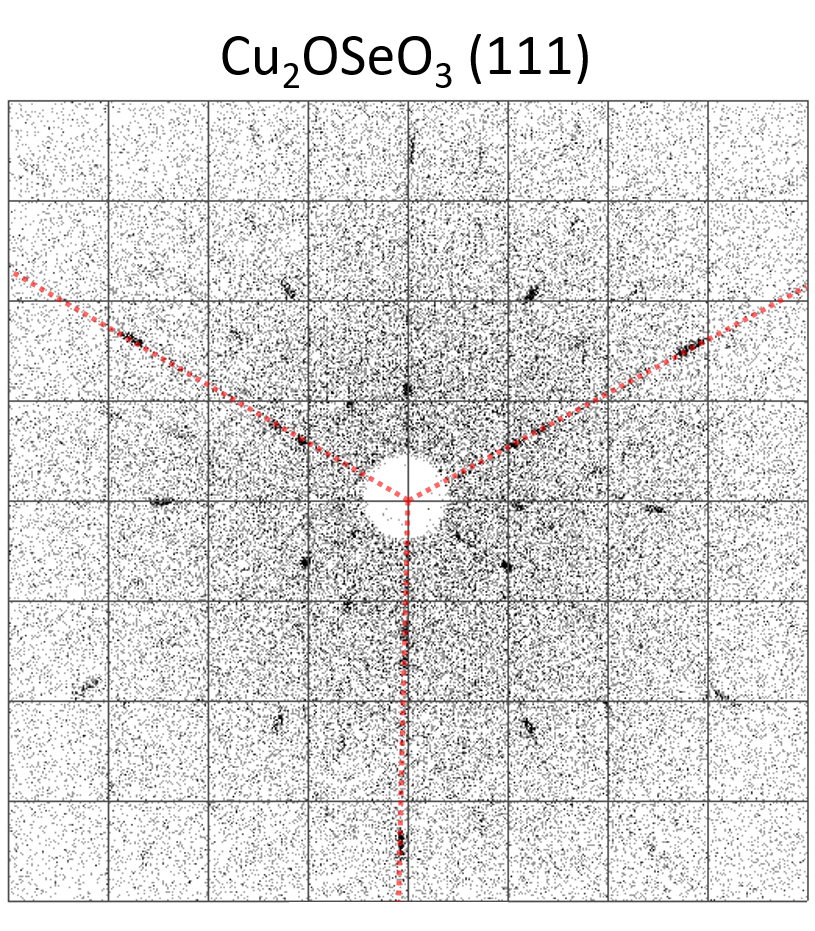}
	\caption{Laue pattern from back-reflection topograph of the (111)-oriented CSO crystal.}
	\label{fig:laue}
\end{figure}

In Fig.~\ref{fig:laue} the back-reflection Laue image of the (111)-oriented CSO crystal is shown. The red dashed lines indicate the threefold symmetry of the Laue pattern.

\section{Sample preparation}
\label{sec:sample}

Before sputtering the Py layer on top of the CSO crystal, we polished its surface and cleaned it in an acetone bath. After sputtering, we annealed the sample in vacuum at $T = 250^{\circ}$ for 5 minutes.
The parameters used during the sputter process are listed in table \ref{tab:torques}.

\begin{table}[htb!]
	\caption{Sputtering parameters of the NiFe grown on CSO. The sample was grown at a pressure $p = 5 \times 10^{-3}$ mbar.}
	\begin{ruledtabular}	
		\begin{tabular}{ccccc}
			Target material  & power (W) & rate ($\si{\angstrom} /$s) & time (s) & type\\
			\hline
			NiFe & 15 & 1.7 & 235 & face-to-face \\
		\end{tabular}
		\label{tab:torques}
	\end{ruledtabular}
\end{table} 

\section{Dependence of the hybrid mode on the external magnetic field and the temperature}
\label{sec:colorplot}

\begin{figure}[htb!]
	\includegraphics{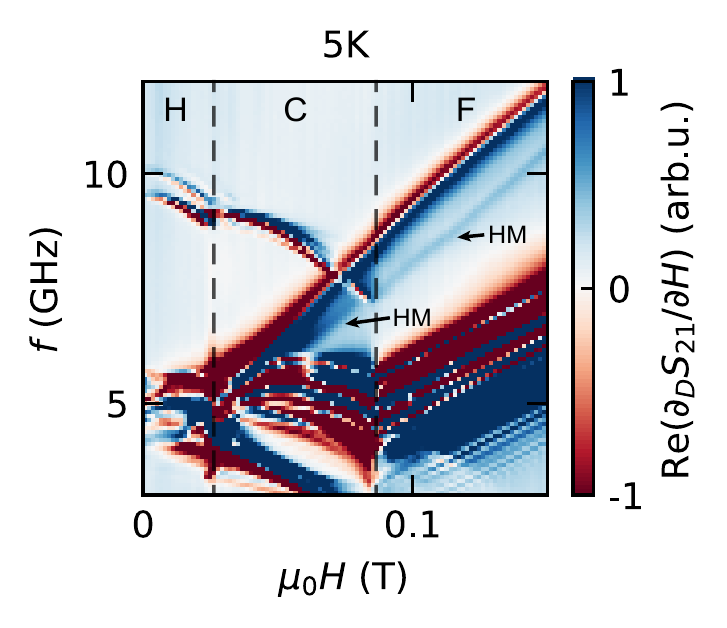}
	\caption{$\mathrm{Re}(\mathrm{\partial}_{\mathrm{D}} S_{\mathrm{21}}/\mathrm{\partial}\mathrm{H})$ at 5~K as a function of the frequency $f$ and the external magnetic field $H$. $H$ is applied along the $\mathbf{x}$-axis ($\phi_H~=~0$). Approximate phase transitions of the helical (H), conical (C), and ferrimagnetic (F) phases are marked by the dashed vertical lines. The hybrid mode (HM) appears in the conical and ferrimagnetic phase of the CSO.}
	\label{fig:conical}
\end{figure}

In Fig.~\ref{fig:conical} the background-corrected field-derivative [\onlinecite{MaierFlaig.2018}] of the VNA transmission spectra $\mathrm{Re}(\mathrm{\partial}_{\mathrm{D}} S_{\mathrm{21}}/\mathrm{\partial}\mathrm{H})$ at 5~K is shown as a function of the frequency $f$ and the external magnetic field amplitude $H$. The external field $\mathbf{H}$ is applied along the $\mathbf{x}$-axis ($\phi_H~=~0$). 
The helical (H), conical (C), and ferrimagnetic (F) phases of the CSO can be observed. The hybrid mode (HM) clearly appears in the conical and ferrimagnetic phases of the CSO.

\begin{figure*}[htb!]
	\includegraphics[width=140mm]{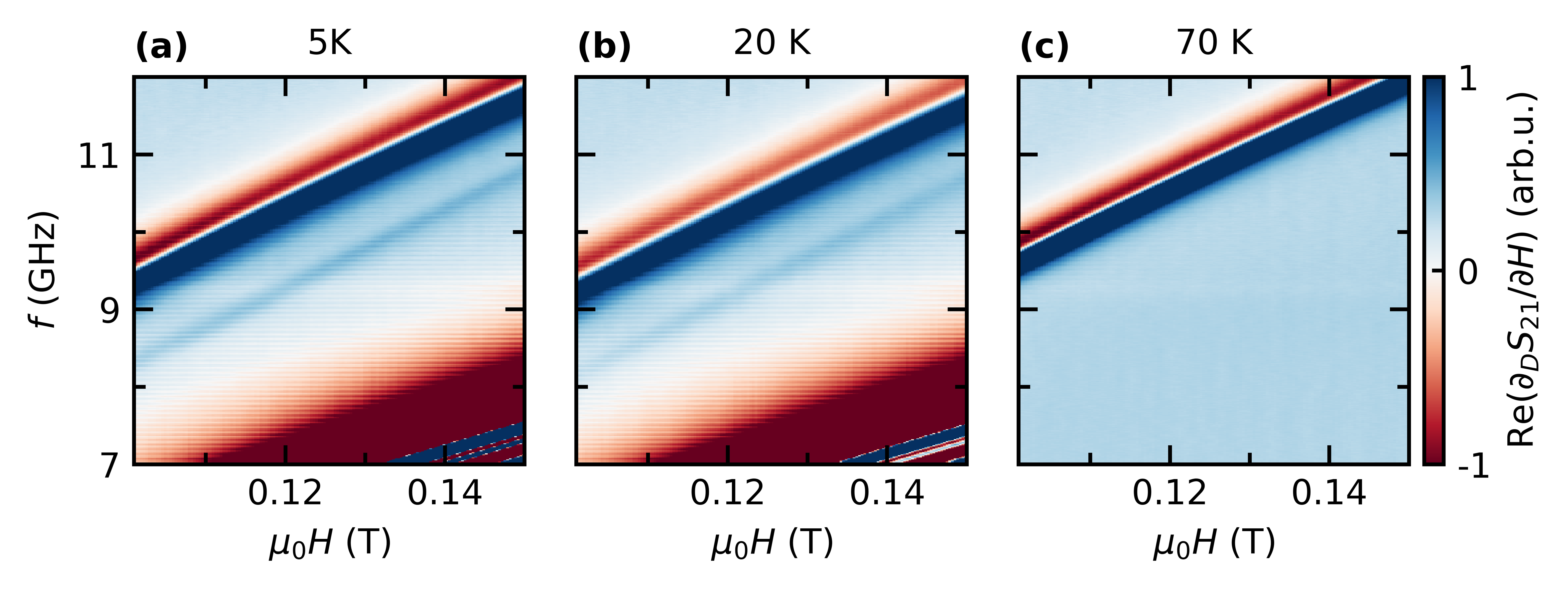}
	\caption{Background-corrected field-derivative \cite{MaierFlaig.2018} of the VNA transmission spectra $\mathrm{Re}(\mathrm{\partial}_{\mathrm{D}} S_{\mathrm{21}}/\mathrm{\partial}\mathrm{H})$ at different temperatures as a function of the frequency $f$ and the external magnetic field $H$ in the ferrimagnetic phase of the CSO. The external field is applied along the $\mathbf{x}$-axis. (a) and (b) The Py FMR mode, the HM mode and the CSO modes appear for temperatures beneath the critical temperature $T_c$ of the CSO. (c) For temperatures higher than $T_c$ the HM does not appear as CSO is not in the ferrimagnetic phase any more.}
	\label{fig:temperature}
\end{figure*}

In Fig.~\ref{fig:temperature} $\mathrm{Re}(\mathrm{\partial}_{\mathrm{D}} S_{\mathrm{21}}/\mathrm{\partial}\mathrm{H})$ for different temperatures is shown as a function of the frequency $f$ and the external field $H$ in the ferrimagnetic phase of the CSO. Fig.~\ref{fig:temperature}~(a) shows the transmission spectra at 5~K. The variation of the HM mode frequency with $\mathbf{H}$ is parallel to that of the Py FMR mode. It is shifted about 1~GHz to lower frequencies than the Py FMR mode. The low frequency modes are attributed to magnetostatic modes of the CSO crystal. Fig.~\ref{fig:temperature}~(b) shows the transmission spectra at 20 K. The HM still appears with a dispersion parallel to the Py FMR mode. Fig.~\ref{fig:temperature}~(c) shows the transmission spectra at 70 K. For temperatures above the critical temperature of the CSO the HM does not appear.\\

\section{Fitting method of the FMR peaks}
\label{sec:fitting}

\begin{figure}[htb!]
	\includegraphics[width=80mm]{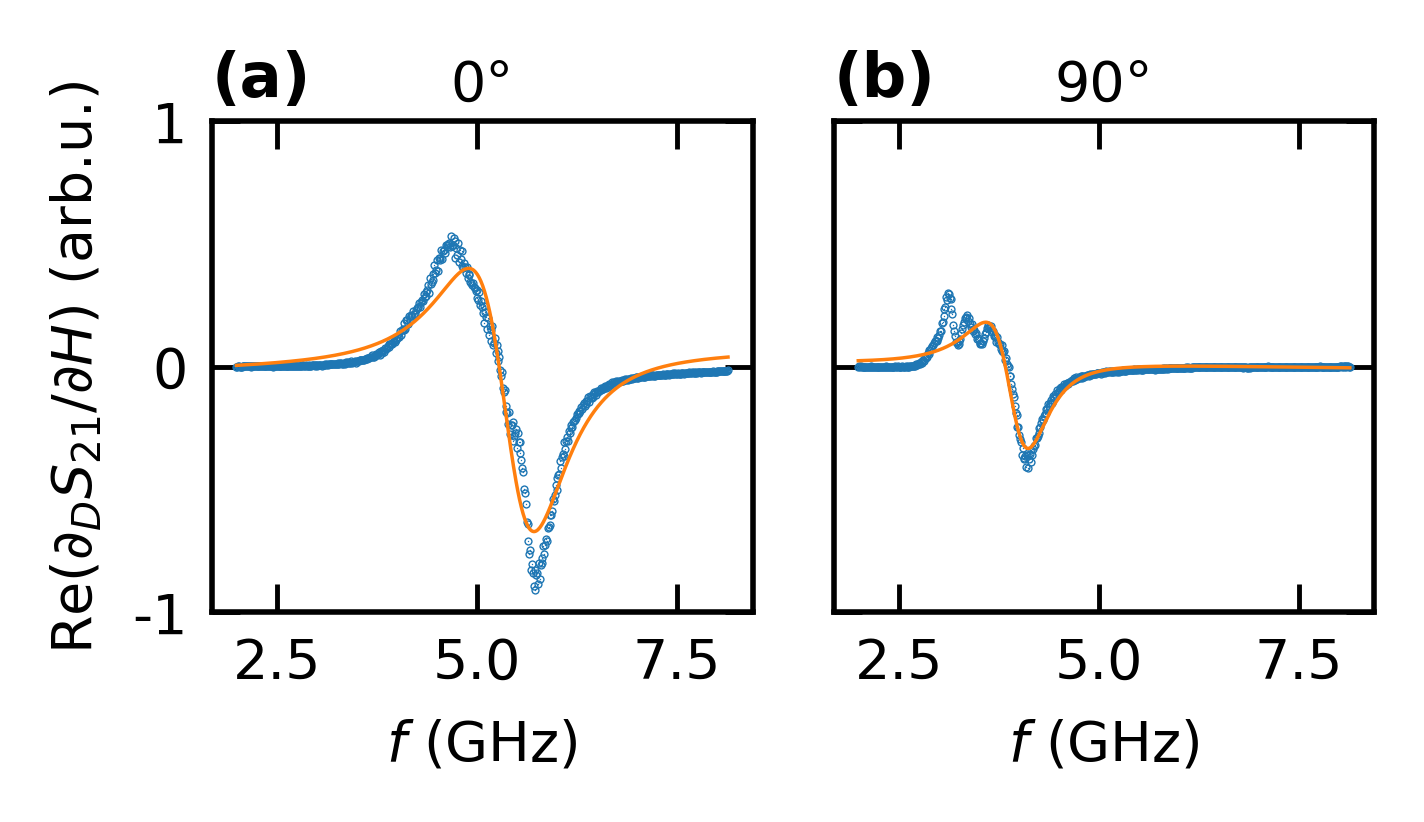}
	\caption{The solid lines (orange) show the fit of the CSO data (blue dots). (a) The external field is applied along the $\mathbf{x}$-axis $(\phi_H~=~0)$° in the Py thin film plane. (b) The external field is applied under an angle $\phi_H~=~90$°.}
	\label{fig:CSOfitting}
\end{figure}

The Py FMR peak and the HM peak were simultaneously fitted in the frequency domain of the background corrected field-derivative of the transmission parameter $S_{21}$ for each fixed external field direction. As a fit model we used the central difference quotient $S_{21}$ given by \cite{MaierFlaig.2018}
\begin{equation}
	d_DS_{21} = i \omega A e^{-i \phi} \frac{\chi(\omega + \Delta \omega_{\pm})-\chi(\omega - \Delta \omega_{\pm})}{2 \Delta \omega_{\pm}} + S_{21_b},
	\label{eq:dS}
\end{equation}
where $S_{21_b}$ corrects for the offset and the slope of the background of the $S_{21}$ signal, $A$ is the modified signal amplitude, which also takes the partial derivative $\frac{\partial \omega}{\mathrm{d}H}$ into account and $\Delta \omega_{\pm} = \Delta H_{\pm} \frac{\partial \omega}{\partial H_{\mathrm{ext}}} \approx \partial H_{\mathrm{ext}} \gamma \mu_0$ is the modulation amplitude with the field step width $ \Delta H_{\pm}$.
The high frequency magnetic susceptibility $\chi$ is given by \cite{MaierFlaig.2018} 
\begin{equation}
	\chi (\omega , H_{\mathrm{ext}}) = \frac{\omega_M (\gamma \mu_0 H_{\mathrm{ext}} + i \Delta \omega)}{(\omega_{\mathrm{res}}(H_{\mathrm{ext}}))^2 - \omega^2 + i\omega \Delta \omega},
\end{equation}
where $\omega_{\mathrm{res}}$ is the resonance frequency, $\Delta \omega$ the full width at half maximum linewidth, and $\omega_M~=~\gamma~\mu_0~M_s$. The ferrimagnetic CSO modes were simultaneously fitted as one mode only with the fit model given in Eq. (\ref{eq:dS}). To determine the boundaries for the frequency range of the CSO modes in dependence of the angle under which the external magnetic field was applied, we only took modes into account with an amplitude bigger than a threshold of about $1/8$ of the maximum CSO resonance amplitude. This was done to avoid designating random noise as part of the resonance frequencies. Two exemplary fits of the CSO modes where the external magnetic field is applied under an angle $\phi_H~=~0$° and $\phi_H~=~90$° with respect to the $\mathbf{x}$-axis are shown in Fig. \ref{fig:CSOfitting}.

\section{Determination of the material parameters}
\label{sec:parameters}

To determine the g-factor $g$ and the saturation magnetization $M_s$ of the Py layer at 5~K, the Py FMR line was fitted by the in-plane Kittel equation
\begin{equation}
	\mu_0 H_{\mathrm{ext}} = -\mu_0 H_{\mathrm{ani}} - \frac{\mu_0M_{s}}{2} + \sqrt{\bigg(\frac{f h}{g \mu_B}\bigg)^2 +\bigg(\frac{\mu_0M_{s}}{2}\bigg)^2 }.
\end{equation}
The damping parameter $\alpha$ was fitted by
\begin{equation}
	\mu_0 H_{\mathrm{ext}} = \mu_0 H_{\mathrm{inh}}+ \frac{4 \pi \alpha \hbar}{g \mu_B}f.
\end{equation}
The thereby obtained values $\mu_0M_{s}~=~1$~T, $\alpha~=~0.01$, $g~=~2.07$ are in good agreement with previous reports \cite{Li.2016}.

\section{Uniaxial anisotropy of the Py resonance frequencies}
\label{sec:uniaxialani}

\begin{figure}[htb!]
	\includegraphics[width=60mm]{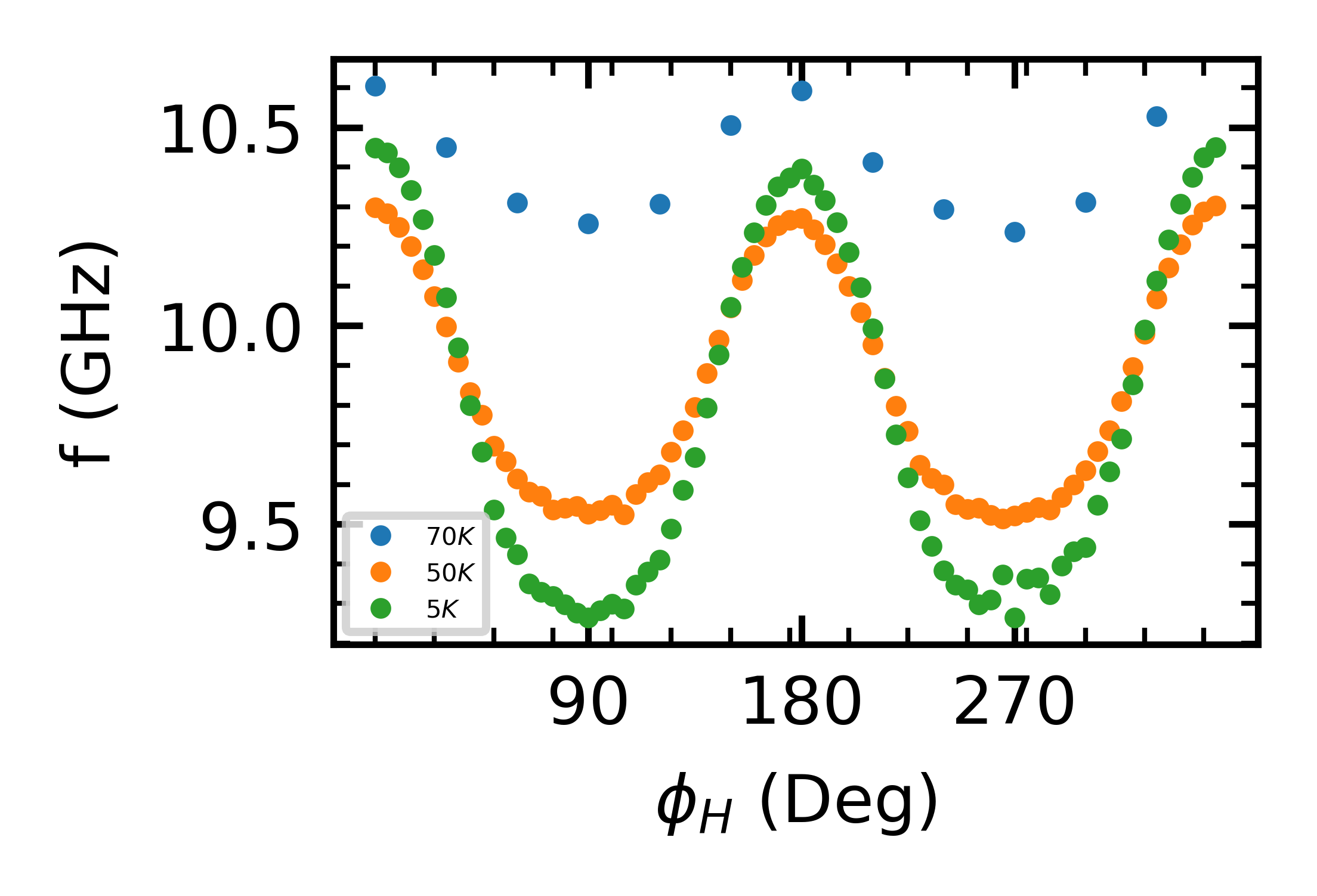}
	\caption{Fitted Py resonance frequencies plotted against the angle under which the external magnetic field is applied at 5~K (green dots), 50~K (orange dots), and 70~K (blue dots). The uniaxial anisotropy is strongest for low temperatures. Above $T_c~=~58.2$~K it decreases considerable.}
	\label{fig:uniani} 
\end{figure}

In Fig. \ref{fig:uniani} the fitted Py resonance frequencies are shown for a 360° angle rotation of the external magnetic field applied in the Py thin film plane at 5~K, 50~K, and 70~K. The uniaxial anisotropy in the Py resonance frequencies decreases considerable for temperatures above $T_c~=~58.2$~K. To estimate the effect of the CSO magnetic stray field on the Py magnetization dynamics we calculated the difference of the  magnetic stray field on top of the CSO crystal for an external field applied in the (111)-plane of the CSO crystal along the long side of the crystal and along the short side of the crystal according to Ref. \onlinecite{EngelHerbert.2005}. This resulted in an anisotropy field $H_{ani}~=~50$~mT at 5~K, which corresponds to a frequency shift of the Py resonance of about 2~GHz when the external magnetic field with a magnitude of 120~mT is rotated from $\phi_H~=~0$° to $\phi_H = 90$°. This is in the order of magnitude of the observed uniaxial anisotropy of the Py resonance frequencies.
Thus, we attribute the observed uniaxial anisotropy of the Py resonance frequencies to the effect of the magnetic stray field coming from the CSO below the Py layer. Above $T_c$ the CSO is not magnetically ordered and the anisotropy in the resonance frequencies decreases a lot. We still observe a maximal frequency shift of the Py resonance frequencies of about 0.36 GHz at 70~K, which corresponds to an anisotropy field of 7.5~mT. This might be caused by the trapped flux in the superconducting coils of the 3D-vector magnet, as approximately the same shift appears at 300~K (data not shown). The setup frequently exhibits trapped flux in the range of 10~mT which substantiates the assumption. Furthermore, small contributions can be induced by the not perfect alignment of the sample in the 3D vectormagnet.

\section{Simulation method}
\label{sec:simulation}

To simulate the resonance frequencies of the Py, CSO, and HM we solve the Landau-Lifshitz-Gilbert (LLG) equation
\begin{eqnarray}
	\frac{\mathrm{d}\mathbf{M}_i}{\mathrm{d}t} =&& - \gamma_i \mathbf{M}_i \times \mu_0 \mathbf{H}_{\mathrm{eff}_i}(\phi_H) + \frac{\alpha_i}{M_{s_i}}\mathbf{M}_i \times \frac{\mathrm{d}\mathbf{M}_i}{\mathrm{d}t}
	\label{eq:LLG}
\end{eqnarray}
in a coordinate system where $\mathbf{x}$ corresponds to the in-plane equilibrium orientation of all the macrospins, $\mathbf{z}$ is the oop direction, and $\mathbf{y}$ the in-plane direction perpendicular to $\mathbf{x}$.
We consider a rotation of the external magnetic field in the film plane with an angle $\phi_H$ to the long side of the CSO crystal (along the CPW center conductor). Furthermore, we assume that the magnetization in the sample is fully aligned with the external magnetic field in equilibrium.
The effective magnetic field $\mathbf{H_{\mathrm{eff_{py}}}}$ of the Py accounts for the external magnetic field $\mathrm{\mathbf{H_{\mathrm{ext}}}}$, the uniaxial anisotropy field $B_{\mathrm{u}}$, the cubic anisotropy field $B_{\mathrm{c}}$ that we assume to exist at the interface to explain the data, and the demagnetization field of a thin film.
\begin{align}
	\begin{split} 
		\mathbf{H_{\mathrm{eff_{py}}}} &=  (\mu_0 \mathrm{H_{ext}}+\mathrm{B_{\mathrm{u}}} \sin{(\phi_H)}^2  + \mathrm{B_{\mathrm{c}}} \sin{(2\phi_H)}^2)\hat{\textbf{x}} \\
		&- \mu_0 m_{z}(t)\hat{\textbf{z}}.
	\end{split}
	\label{eq:Heffpy}
\end{align} 
The effective magnetic field $\mathbf{H_{\mathrm{eff_{HM}}}}$ of the HM accounts for the external magnetic field $\mathrm{\mathbf{H_{\mathrm{ext}}}}$, the cubic anisotropy field $B_{\mathrm{c}}$, and the demagnetization field of a thin film.
\begin{align}
	\begin{split} 
		\mathbf{H_{\mathrm{eff_{HM}}}} &=  (\mu_0 \mathrm{H_{ext}} + \mathrm{B_{\mathrm{c}}} \sin{(2\phi_H)}^2)\hat{\textbf{x}} \\
		&- \mu_0 m_{z}(t)\hat{\textbf{z}}.
	\end{split}
	\label{eq:HeffHM}
\end{align} 
The effective magnetic field $\mathbf{H_{\mathrm{eff_{CSO}}}}$ of the CSO accounts for the external magnetic field $\mathrm{\mathbf{H_{\mathrm{ext}}}}$ and the demagnetization field of a bulk crystal.
\begin{eqnarray}
	\mathbf{H_{\mathrm{eff_{CSO}}}} =  \mu_0 \mathrm{H_{ext}} \hat{\textbf{x}} - \mu_0  \left( \begin{array}{c} N_x(\phi_H) \cdot M_s\\ N_y(\phi_H) \cdot m_{y_j}(t)\\  N_z\cdot m_{z_j}(t) \\\end{array}\right),
	\label{eq:HeffCSO} 
\end{eqnarray}
with the demagnetization factors $N_x(\phi_H) = N_{x_0}\cos(\phi_H)^2 + N_{y_0} \sin(\phi_H)^2$ and $N_y(\phi_H) = N_{y_0}\cos(\phi_H)^2 + N_{x_0} \sin(\phi_H)^2$, with $N_{x_0}~=~0.145$, $N_{y_0}~=~0.285$, and $N_z~=~0.570$, which we calculated from the crystal geometry according to Ref. \onlinecite{Osborn.1945}. 
Furthermore, we use for the dynamic magnetization the ansatz $m_{y_i,z_i}(t)~=~m_{y_i,z_i}~\cdot~e^{-i \omega t}$.
The temperature independent material parameters for the simulation of the Py are $\alpha~=~0.01$, $g~=~2.07$, as extracted in Supplemental Material \ref{sec:parameters}. For CSO we use  $\alpha~=~0.003$, $g~=~2$, which are in good agreement with previous reports \cite{Weiler.2017,Li.2016,Belesi.2011}. For the HM we use a saturation magnetization which is lower than the Py saturation magnetization $M_{s_{\mathrm{HM}}}~=~0.75~M_{s_{\mathrm{Py}}}$. We motivate this assumption by the lower saturation magnetization of the CSO as well as the findings in Refs [\onlinecite{Hellman.2017}] and [\onlinecite{Vaz.2008}], where it was shown that the saturation magnetization at the interface of two magnetic materials can be varied due to interfacial effects.
Furthermore, the temperature dependent parameters used in our numerical calculations are given in Tab. \ref{tab:params}.
\begin{table}[htb!]
	\caption{Temperature dependent parameter sets used for the numerical solving of Eq. (\ref{eq:LLG}) at $T~=~5$K and $T~=~50$K.}
	\begin{ruledtabular}
		\begin{tabular}{cccc}
			Layer  & $\mathrm{B_{\mathrm{u}}}$ (mT) & $\mathrm{B_{\mathrm{c}}}$ (mT) & $\mathrm{M_s}$ (A/m) \\
			\hline
			NiFe & -23 (T~=~5~K) & -3 (T~=~5~K) & 1/$\mu_0$ (T~=~5~K) \\
			\hline
			NiFe & -16 (T~=~50~K) & -1.5 (T~=~50~K) & 0.98/$\mu_0$ (T~=~50~K) \\
			\hline
			HM & - & -6 (T~=~5~K) & 0.755/$\mu_0$ (T~=~5~K) \\
			\hline
			HM & - & -3.4 (T~=~50~K) & 0.82/$\mu_0$ (T~=~50~K) \\
			\hline
			CSO & - & - & 0.4/$\mu_0$ (T~=~5~K) \\
			\hline
			CSO & - & - & 0.21/$\mu_0$ (T~=~50~K) \\
		\end{tabular}
		\label{tab:params}
	\end{ruledtabular}
\end{table}

\bibliography{Quellen}

\end{document}